# Emergence of equilibrium thermodynamic properties in quantum pure states II. Analysis of a spin model system.

**Barbara Fresch**[a] , **Giorgio J. Moro**[b]

*Department of Chemical Science, University of Padova, Via Marzolo 1, 35131 Padova – Italy*

**Abstract**

A system composed of identical spins and described by a quantum mechanical pure state is analyzed within the statistical framework presented in the part I of this work. We explicitly derive the typical values of the entropy, of the energy, and of the equilibrium reduced density matrix of a subsystem for the two different statistics introduced in part I. In order to analyze their consistency with thermodynamics these quantities of interest are evaluated in the limit of large number of components of the isolated system. The main results can be summarized as follows: typical values of the entropy and of the equilibrium reduced density matrix as functions of the internal energy in the Fixed Expectation Energy Ensemble do not satisfy the requirement of thermodynamics. On the contrary, the thermodynamical description is recovered from the Random Pure State Ensemble (RPSE), provided that one considers systems large enough. The thermodynamic limit of the considered properties for the spin system reveals a number of important features. Firstly canonical statistics (and thus canonical typicality as long as the fluctuations around the average value are small) emerges without the need of assuming the microcanonical space for the global pure state. Moreover, we rigorously prove i) the equivalence of the "global temperature", derived from the entropy equation of state, with the "local temperature" determining the canonical state of the subsystems; and ii) the equivalence between the RPSE typical entropy and the canonical entropy for the overall system.

[a] barbara.fresch@unipd.it

[b] giorgio.moro@unipd.it

## I. INTRODUCTION

In this paper we shall apply the statistical theory developed in the previous paper *Emergence of equilibrium thermodynamic properties in quantum pure states I*, (in the following indicated as 'I'), to a model system composed of $n$ spins with quantum spin number $J=1$. Spin systems are convenient models for investigations of quantum statistical behaviour, since one has to consider a finite dimensional Hilbert space. The numerical calculations of the energy spectra and the time evolution of arbitrary initial states can be performed to machine precision without introducing any artificial truncation of the Hilbert space. Furthermore this kind of model systems is the subject of a continuously increasing attention either from a theoretical [1] as well as experimental perspective [2] because it represents the natural test bed for quantum information protocols.

The purpose of this second part of the work is twofold. Firstly we shall verify the existence and the calculability of typical values of some functions of interest in the ensemble of pure states defined in I. In particular, we shall focus on the internal energy and the entropy, as the fundamental entities of thermodynamics; moreover we shall also study the equilibrium state of a single component (subsystem) of the entire isolated system. We anticipate that such typical values are well defined in both our ensembles. Furthermore, by employing the probability distributions on populations derived in paper I we shall derive explicitly the typical value dependence on the type of ensemble, i.e. on the relevant constraints which specify the space of the pure states of the isolated system. Notice that explicit results on the reduced density matrix of a subsystem has been previously derived [10], in the asymptotic limit of large isolated system, only by considering the particular space of pure states which lie in a narrow energy window traditionally associated to the microcanonical condition.

The analysis here presented shows that the behaviour of the typical values of the considered thermodynamic functions and the typical state of a subsystem is indeed remarkably different for the two choices of the population statistics, i.e. for the Fixed Expectation Energy Ensemble (FEEE) and the Random Pure State Ensemble (RPSE). The congruence between the behaviour of typical values and standard thermodynamics provides the basis for the choice of the appropriate ensemble. To this aim the asymptotic form for $n \to \infty$ are derived for the functions of interest.

Secondly, by studying the ensemble distribution of the functions of interest one can quantify their typicality. That is, while univocally defined value of the thermodynamic functions of the single pure state are recovered only in the thermodynamic limit, i.e. in the limit of infinite size of the system, in numerical experiments one needs to deal with finite systems. Thus the following question arises: down to which size of the system can the standard thermodynamic concepts meaningfully be defined and employed? Since in our treatment thermodynamic properties are emergent properties, one can quantitatively calculate the deviation of the properties of a system from their asymptotic typical values due to the finiteness of the considered system. To this aim we shall always show together with the derived asymptotic values, the results of calculations for finite size system with varying dimensions.

For the purposes of the present paper, which is confined to the analysis of the equilibrium properties, an ideal system will be considered. The absence of the interaction term in the Hamiltonian has to be intended as an approximation that allows a simple evaluation of the energy spectrum of the system. Indeed the presence of weak interactions amongst the components (spins) modifies the energy spectrum only slightly and thus it does not affect the statistical characterization of the equilibrium state that we shall consider. It should be clear that the interactions imply the development of correlations among the components of the global system leading to the necessity of considering only the entire composite system as describable by a pure state. Of course, the presence of the interactions also plays an essential role in addressing the issue of relaxation toward the equilibrium state but we do not consider this problem here.

The paper is organized as follows. In the next Section the model spin system is introduced, and its energy spectrum is characterized. In Section III we report the results of the FEEE statistics. In the following Section the same analysis is done for the RPSE. In the final Section, together with general conclusions, we discuss the critical issues deriving from the present work.

## II. THE MODEL

In order to investigate the emergence of thermodynamic functions in a statistical ensemble, we shall consider systems composed of $n$ identical components (or subunits), each of them described in a finite dimensional Hilbert space. Then, the Hilbert space of the total system is the tensor product of the Hilbert spaces of the components

$$\mathcal{H} = \mathcal{H}_1 \otimes \mathcal{H}_1 .... \otimes \mathcal{H}_n \tag{1}$$

A prototype of this kind of composite systems is an ensemble of spins. In the following we shall specifically consider systems of $n$ identical spins having $J = 1$ as the spin quantum number, whose total Hilbert space is spanned by vectors of dimension $N = 3^n$. Since we consider an ideal system (i.e., no interactions between spins), the total Hamiltonian includes only the Hamiltonians $H^{(j)}$ of each spin $j$

$$H = \sum_{j=1}^{n} H^{(j)} \tag{2}$$

For the sake of simplicity, we model the spin Hamiltonians according to Zeeman interactions along the $z$ axis with the same transition frequency $\omega_0$:

$$H^{(j)} = \hbar \omega_0 S_z^{(j)} \tag{3}$$

$S_z^{(j)}$ being the $z$ component of the $j$-the spin operator. Therefore the eigenstates $|m^{(j)}\rangle$ of $S_z^{(j)}$ for $m^{(j)} = -1, 0, 1$ allow us to specify the energy eigenstates $|M\rangle \equiv |m^{(1)}\rangle |m^{(2)}\rangle \cdots |m^{(n)}\rangle$ for the overall system

$$H|M\rangle = E_M |M\rangle \tag{4}$$

with energy eigenvalues

$$E_M / \hbar \omega_0 = \sum_{j=1}^{n} m^{(j)} \tag{5}$$

where $M$ denotes the sequence of spin components: $M \equiv (m^{(1)}, m^{(2)}, \cdots, m^{(n)})$. Notice that the eigenenergies are within the range $-n \leq E_M / \hbar \omega_0 \leq n$. For later elaborations, it is convenient to order the energy eigenvalues in magnitude according to an index $k$ such that $E_k \leq E_{k+1}$ for $k = 1, 2, \cdots, N$.

The energy spectrum, because of the equivalence of the spins, is highly degenerate. Given the set of occupation numbers $(i_{-1}, i_0, i_1)$ of an eigenstate for the spins with components $m = -1$, $m = 0$ and $m = 1$, respectively, the degeneracy of the eigenstate is given by $n!/(i_{-1}! i_0! i_1!)$. Because of the constraint $i_{-1} + i_0 + i_1 = n$, only two occupation numbers are independent, and they are conveniently chosen as $i_0$ and $i \equiv i_1 - i_{-1}$. In this way the energy

$$E / \hbar \omega_0 = i \tag{6}$$

results to be independent of $i_0$ occupation number. Notice that, by taking into account that $i_1 = (n - i_0 + i)/2$ and $i_{-1} = (n - i_0 - i)/2$ should not be negative, one derives for the two independent parameters $(i, i_0)$ the domain which is represented in Fig. 1. The degeneracy $D(n, i)$ of the energy eigenvalue $i\hbar \omega_0$ of the $n$ spin system for $i = -n, -n+1, \cdots, n$ is derived by adding the degeneracy for all the possible values of $i_0$, that is

$$D(n,i) = \sum_{i_0 = 0}^{\min(n-i, n+i)} \frac{n!}{[(n - i_0 + i)/2]! i_0! [(n - i_0 - i)/2]!} \tag{7}$$

Of course, the dimension $N = 3^n$ of the overall Hilbert space is recovered from the sum of the degeneracies of all the energy eigenvalues

$$\sum_{i=-n}^{n} D(n,i) = 3^n \tag{8}$$

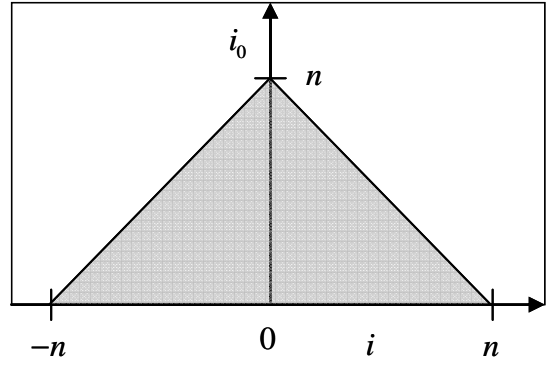

**Fig. 1:** Domain for the independent occupation numbers $(i, i_0)$ represented as the grey area.

In the following analysis, often we need to evaluate quantities like $\sum_{k=1}^{N} f(E_k)$ for a given function $f(E)$. Their exact calculations are efficiently performed by employing the degeneracy eq.(7) of the energy eigenvalues

$$\sum_{k=1}^{N} f(E_k) = \sum_{i=-n}^{n} D(n,i) f(i\hbar\omega_0) \qquad (9)$$

On the other hand, in order to derive their asymptotic expressions in the limit of large system size, i.e., for $n \to \infty$, it is convenient to introduce the density of states

$$g(E) = \sum_{k=1}^{N} \delta(E - E_k) \qquad (10)$$

which allows their calculation by integration on the energy variable:

$$\sum_{k=1}^{N} f(E_k) = \int dE \, f(E) g(E) \qquad (11)$$

A simple approximation of the density of states is readily obtained by considering that the total energy eq.(5) can be interpreted as the sum of $n$ random (discrete) variables $m^{(j)}$. Then, from the Central Limit Theorem [3] the density of states is fairly well approximated by a Gaussian function

$$g(E) = \frac{N}{\sqrt{2\pi\sigma^2}} e^{-\frac{(E-\bar{E})^2}{2\sigma^2}} \qquad (12)$$

whose parameters $\bar{E}$ and $\sigma$ are derived on the basis of the properties of the spectra of the single components:

$$\bar{E}/\hbar\omega_0 = \sum_{j=1}^{n} \overline{m^{(j)}} = 0 \qquad \sigma^2/(\hbar\omega_0)^2 = \sum_{j=1}^{n} \overline{(m^{(j)} - \overline{m^{(j)}})^2} = 2n/3 \qquad (13)$$

As an example, in Fig.2 we compare the Gaussian density of states with the energy eigenvalue degeneracy for the case of ten ($n = 10$) spins. Notice that by considering the energy variable scaled according to its half-domain width $n\hbar\omega_0$, one recovers a narrowing of the density of states with the increase of the number of spin, since $\sigma/n\hbar\omega_0 = \sqrt{2/3n}$. Therefore, in the asymptotic limit of a large number of spins, the density of states can be approximated by a Dirac delta function

$$n \to \infty: \quad g(E) = N\delta(E) \tag{14}$$

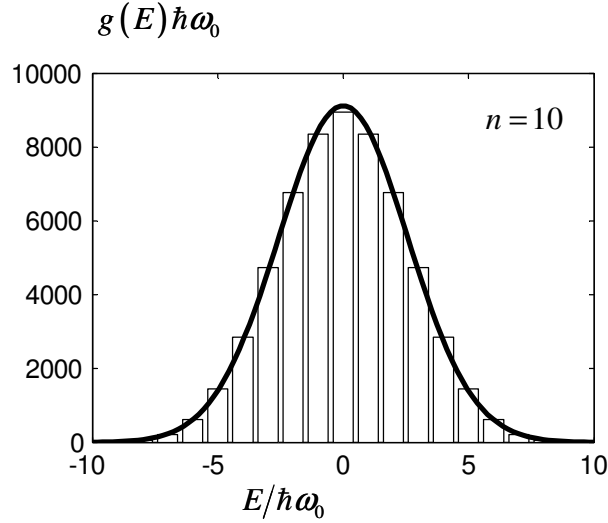

**Fig. 2: Gaussian density of states of ten identical $J=1$ spins (solid line). The histogram shows the degeneracy calculated according to eq.** (7).

It should be also mentioned that the Gaussian profile of the density of states does not depend on the peculiar form of the energy spectrum of each subsystem, as long as it is bounded. This because the overall energy density of composite systems results from the convolution of the single component energy density and, if the energy spectrum of each component is bounded, then, by virtue of the Central Limit Theorem, the total energy density can be in general well approximated with a Gaussian function [4].

## III. THE FIXED EXPECTATION ENERGY ENSEMBLE (FEEE)

As reported in detail in Section II-A of I the pure state of the isolated composite system is characterized by a fixed set of populations $P = (P_1,...,P_N)$ and a set of phases which results homogenously distributed as a consequence of the Schrödinger evolution. Thus, equilibrium properties, which are defined as asymptotic time averages, are explicit functions of the population set.

In this section we shall consider, within the FEEE, the emergence of typical values of the entropy

$$S = -k_B \sum_{k=1}^{N} P_k \log P_k \tag{15}$$

and of the equilibrium reduced density matrix $\bar{\mu}$ of a spin. If typical values of these quantities effectively emerge, then we can investigate their functional dependence on the internal energy $U$ which, in this ensemble, coincides with the expectation energy: $U = E = \sum_{k=1}^{N} P_k E_k$.

The analysis is greatly simplified by the use of the approximate form of the FEEE distribution derived in Ref. [5] which assures good estimates of the ensemble averages. The main advantage of such an approximate distribution derives from the factorization of the population distribution into single variable distributions, as recalled by eq. (30) of I

$$p_{FEEE}(P) \simeq \prod_{k=1}^{N} w^{(k)}(P_k) \qquad (16)$$

where for any variable $P_k$, the domain $[0, \infty)$ is considered. The single variable distributions read

$$w^{(k)}(P_k) = \frac{1}{\langle P_k \rangle} e^{-P_k / \langle P_k \rangle} \qquad k \neq 1$$

$$w^{(1)}(P_1) = \delta(P_1 - \langle P_1 \rangle) \qquad (17)$$

where the average of the first population is

$$\langle P_1 \rangle = 1 - \frac{U - E_1}{N-1} \sum_{k=2}^{N} \frac{1}{E_k - E_1} \qquad (18)$$

while the average of the other populations are given as

$$\langle P_k \rangle = \frac{E}{(N-1) E_k} \qquad (19)$$

In Ref. 5 (see in particular Fig. 5 therein) we have reported calculations for spin systems clearly showing the emergence of entropy typicality for an increasing number $n$ of spins. This happens even if the populations are broadly distributed, like in the distributions $w^{(k)}$ of eq. (17) for $k \neq 1$. The entropy typicality can be quantified by computing its variance $\sigma_S$ within the ensemble, while its typical value can be identified with its average $\langle S \rangle$, both quantities being easily evaluated by means of the approximate distribution eq. (17). For the average we obtain

$$\langle S \rangle = \int dP S(P) p_{FEEE}(P) = -k_B \sum_{k=1}^{N} \langle P_k \rangle \ln \langle P_k \rangle - (1-\gamma)(1 - \langle P_1 \rangle) =$$
$$\simeq (U - E_1) k_B \frac{\ln N}{N} \sum_{k=2}^{N} \frac{1}{E_k - E_1} \qquad (20)$$

where $\gamma \cong 0.5772$ is the Euler constant and the last equation is derived by retaining only the terms of leading order with respect to $N$. By applying the same procedure to the variance, and again by retaining only the leading order terms in $N$, we obtain

$$\sigma_S^2 = \langle S^2 \rangle - \langle S \rangle^2 \simeq (U - E_1)^2 k_B^2 \left( \frac{\ln N}{N} \right)^2 \sum_{k=2}^{N} \frac{1}{(E_k - E_1)^2} \qquad (21)$$

Let us now evaluate the average and the variance of the entropy in the limit of a large number $n$ of spins. The summations $\sum_{k=2}^{N} f(E_k)$, with $f(x)=1/x$ for eq. (20) and with $f(x)=1/x^2$ for eq. (21), can be estimated as integrals [6] by introducing the density of the states eq. (12). On the other hand for large $n$, the Dirac delta approximation eq. (14) can be employed for the density of states, so obtaining

$$\sum_{k=2}^{N} f(E_k) \cong N f(0) + R \qquad (22)$$

It should be evident that, since the first term at the r.h.s. is proportional to $N$, the rest $R$ can be safely neglected. In this way both the average entropy and its variance can be easily evaluated in the asymptotic limit of a large number of spins

$$\langle S \rangle / k_B = (U/\hbar\omega_0 + n)\ln 3 \qquad \sigma_S / k_B = \frac{(U/\hbar\omega_0 + n)\ln 3}{\sqrt{N}} \qquad (23)$$

The condition of typicality is readily verified in the limit of large $n$, since the domain of the entropy is $\Delta S/k_B = n \ln 3$ and thus $\sigma_S / \Delta S \propto 1/\sqrt{N}$.

In figure 3 the FEEE average entropy per spin, evaluated according to eq. (20), is reported as a function of the expectation energy per spin for systems with different numbers of components, $n = 5, 10, 50$, together with the asymptotic formula eq. (23). The convergence to the asymptotic profile as the number of spins increases is evident. Therefore, for macroscopic systems ($n \to \infty$) we recover from FEEE an extensive entropy (that is, $\langle S \rangle \propto n$ for a given $U/n$), but that depends linearly on the internal energy $U$. This, from $1/T = d\langle S \rangle / dU$, implies that the macroscopic system has always the temperature $T = \hbar\omega_0 / k_B \ln 3$ whatever is the internal energy. The impossibility of changing the temperature of the system is evidently a nonsense from the thermodynamic point of view. In conclusion, even if FEEE provides a good statistics for the entropy assuring typicality for large numbers of spins, such a statistical ensemble has to be excluded because it cannot reproduce the thermodynamic behaviour of macroscopic systems.

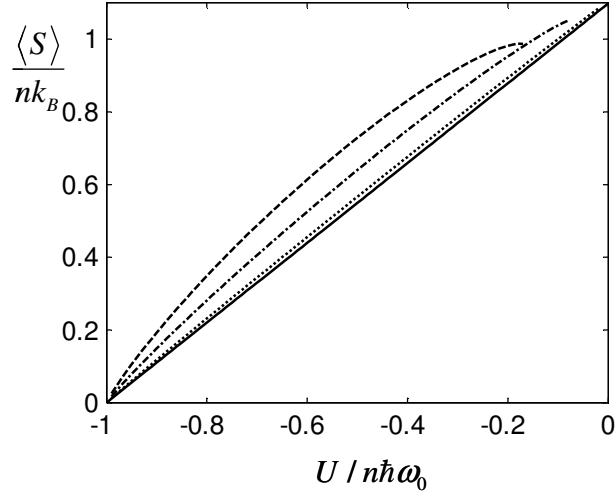

**Fig. 3: Average FEEE entropy per spin as a function of the internal energy per spin for a system composed of $n$ three levels systems. The continuous line represents the asymptotic $n \to \infty$ behavior described by eq.(23), while the other lines are the results of calculation according to eq. (20) for finite systems: $n=5$ (dashed line), $n=10$ (dash-dotted line), $n=50$ (dotted line).**

There is a further motivation for excluding the FEEE model, because it does not assure the canonical form (see eq. (38) of I) for the equilibrium reduced density matrix of a spin considered as a subsystem ($S$) of the isolated sample, with the remaining $n-1$ spins that play the role of environment ($E$). For the ideal system here considered the reduced density matrix is diagonal in the eigenbasis of the single system Hamiltonian, eq. (3), and its elements are determined by linear combinations of the global populations, that is

$$\bar{\mu} = \mathrm{Tr}_E\{\bar{\rho}\} = \sum_s \sum_e P_{se} |s\rangle\langle s| \qquad (24)$$

where $s = 0, \pm 1$ and each population $P_{se}$ corresponds to an energy level of the overall system which is expressed as the sum of the subsystem energy ($E_s^S$) and of the environment energy ($E_e^E$)

$$E_{se} = E_s^S + E_e^E \qquad (25)$$

The typicality of equilibrium reduced density matrix is easily established by invoking the Central Limit Theorem for the sum eq. (24) of stochastic variables exponentially distributed according to eq. (17). Indeed one can easily shows that in the case of $n \gg 1$ the FEEE average $\langle \bar{\mu}_{ss} \rangle$ of an element of the reduced density matrix does not depend on the total number of states $N$, while the corresponding variance scales with the inverse square root of the number of states, $\sigma_{\bar{\mu}_{ss}} \propto N^{-1/2}$.

Let us specify the canonical form of the typical values of the reduced density matrix as

$$\langle \bar{\mu}_{s,s} \rangle = \frac{\exp(-s\hbar\omega_0 / k_B T)}{Q(T)} \qquad (26)$$

where the temperature dependent partition function is given as $Q(T)=\sum_s \exp(-s\hbar\omega_0/k_B T)$. In eq. (26) the temperature has to be considered as an unknown parameter, as long as the system entropy does not provide for it a meaningful value. Therefore, in order to test the validity of the canonical form eq. (26), it is convenient to eliminate the partition function by considering the two ratios $\langle\bar{\mu}_{11}\rangle/\langle\bar{\mu}_{00}\rangle$ and $\langle\bar{\mu}_{00}\rangle/\langle\bar{\mu}_{-1-1}\rangle$ which both should be equal to $\exp(-\hbar\omega_0/k_B T)$. Thus, we introduce the following parameter

$$R = \frac{\langle\bar{\mu}_{11}\rangle\langle\bar{\mu}_{-1-1}\rangle}{\langle\bar{\mu}_{00}\rangle^2} - 1 \tag{27}$$

as a convenient measure of the deviations of the reduced density matrix from the canonical form eq. (26), since $R$ would be null independently of the temperature if the typical values of the reduced density matrix are described by eq.(26).

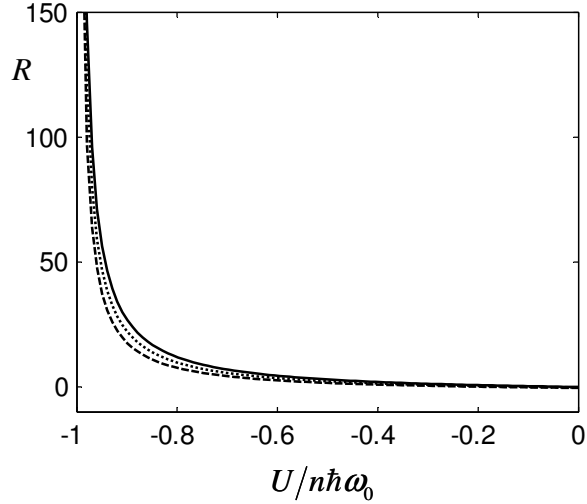

**Fig. 4: Parameter $R$ eq. (27) which quantifies the deviations from the canonical form of the reduced density matrix as a function of the internal energy. The dashed and the dotted lines refer to a system with $n=5$ and with $n=10$ spins, while the continuous line represents the asymptotic result eq. (29) for a large number of spins.**

In Fig. 4 we have reported the values of this parameter as a function of the internal energy for two finite systems with $n=5$ and $n=10$ spins. The typical values of the reduced density matrix elements have been computed according to eq. (24) by inserting average FEEE populations evaluated according to eqs. (18) and (19):

$$\langle\bar{\mu}_{-1-1}\rangle = 1 - \frac{U-E_1}{N-1}\sum_{s\neq-1}\sum_e \frac{1}{E_s^S + E_e^E - E_1}$$
$$s\neq -1:\ \langle\bar{\mu}_{ss}\rangle = \frac{U-E_1}{N-1}\sum_e \frac{1}{E_s^S + E_e^E - E_1} \tag{28}$$

The data displayed in Fig. 4 evidence large deviations from the canonical form of the reduced density matrix. These deviations, however, might be attributed to the finiteness of the considered systems. Indeed, canonical forms should be recovered only in the limit of an infinitely large system, since only in this case the environment behaves like a truly thermal bath constraining the temperature of the subsystem. This calls for the evaluation of the typical reduced density matrix in the asymptotic limit $n \to \infty$.

In strict analogy with the previous analysis of the average entropy, eq. (20), we can approximate the summation on the environment energies according to density of states of $(n-1)$ spins. By invoking again the property that such a density of states is nearly proportional to a Dirac delta function, we obtain the following relations for the elements of the typical reduced density matrix in the FEEE

$$\langle \bar{\mu}_{-1,-1} \rangle = 1 - \frac{U/\hbar\omega_0 + n}{N-1} \sum_{s \neq -1} \frac{N/3}{s+n} \simeq \frac{1 - 2U/n\hbar\omega_0}{3}$$

$$s \neq -1: \quad \langle \bar{\mu}_{ss} \rangle = \frac{U/\hbar\omega_0 + n}{N-1} \frac{N/3}{s+n} \simeq \frac{U/n\hbar\omega_0 + 1}{3}$$

(29)

with the limit $n \to \infty$ leading to the r.h.s of both equations. These asymptotic forms imply, even in the limit $n \to \infty$, non vanishing values of $R$, eq.(27)

$$R = -\frac{3U/n\hbar\omega_0}{U/n\hbar\omega_0 + 1}$$

(30)

which is also displayed in Fig. 4. Recently, the statistics of the FEEE has also been analysed by B. F. Fine in Ref. [7]. The results presented in that work on the statistics of the populations in the FEEE ensemble are in agreement with the results obtained by the authors of the present paper in Ref [5]. Indeed, the asymptotic form of the reduced density matrix for the FEEE, eqs. (29), are equivalent to the application of the formulae (73) and (74) of Ref. [7] to the spin system considered here.

Thus, FEEE generates reduced density matrices which do not have a canonical form. It should be pointed out that such a conclusion can be reached only by considering subsystems with three or more states. If, on the contrary, one examines subsystems with two states, say $J = 1/2$ spin systems, their equilibrium reduced density matrix is specified by one independent parameter only and the temperature can be always derived according to eq. (26). Thus, it can be always written in the canonical form, eq. (26), independently of the thermal properties of the bath.

The conclusion of the analysis presented in this section is thus the following: even if there exists typical values of the thermodynamic functions in the Fixed Expectation Energy Ensemble, these values do not have a well defined thermodynamic behavior. This is certainly true for the entropy which is a property of the whole pure state. Also the equilibrium reduced density matrix of a subsystem displays typicality, but without connections with the standard canonical form. One can suspect that the failure of the FEEE in providing thermodynamically meaningful typical values of the considered quantities for the spin systems arises from the rather special energy spectrum of this model. Indeed, it is generally accepted that while an energy density which monotonically increases with the energy leads to a correct thermodynamical description, different energy density profile, as that characterizing a spin system (see Figure 2), can lead to behaviour which are not consistent with thermodynamics. However it can be shown that the particular energy density profile of spin systems is not the origin of the inconsistency of the FEEE with thermodynamics because the same behaviour of the considered functions (i.e. the entropy and the equilibrium average reduced density matrix) are found for different schemes of the density of states [7,8]. The inconsistency with macroscopic thermodynamics has to be considered as intrinsic to the FEEE statistics on populations.

This conclusion is consistent with the analysis of typicality presented by Popescu et al. [9]. As these authors clearly point out, the problem of the existence of a typical value for the reduced density matrix of a subsystem has to be considered separately from the problem of finding its structure. In general, the typical value, called "generalized canonical state" by these authors, would have a form depending on the type of constraints imposed to the ensemble. Thus our previous analysis can be seen as the derivation of the "generalized canonical state" eq. (28) for the ensemble with the constraint of fixed expectation energy.

## IV. THE RANDOM PURE STATE ENSEMBLE (RPSE)

We recall that the reference space of this ensemble (i.e., the RPSE active space) includes all the wavefunctions which lie in a portion of the overall Hilbert space spanned by the energy eigenvectors which corresponds to the eigenvalues smaller then an arbitrary high energy cut off $E_{\max}$. An important ingredient of our analysis is the dimension $N_{RPSE}(n, E_{\max})$ of the RPSE active space, as a function of the energy cut off and of the number of spins, which can be derived in all generality from the density of states $g(E)$ eq. (10) of the overall Hilbert space

$$N_{RPSE}(n, E_{\max}) = \int_{-\infty}^{E_{\max}} g(E) dE \tag{31}$$

Our analysis requires explicit and simple relations of such a parameter for systems with a large number of spins. This evidently calls for asymptotic approximations of the density of states. The simplest one is certainly the Gaussian density of states eq. (12), to be considered together with the parameters specified by eq. (13), whose substitution into eq. (31) leads to the following result

$$N_{RPSE}(n, E_{\max}) / N = 1 - \mathrm{erfc}\left(E_{\max} / \hbar \omega_0 \sqrt{4n/3}\right) / 2 \tag{32}$$

where $\mathrm{erfc}(x)$ is the complementary error function on the variable $x$.

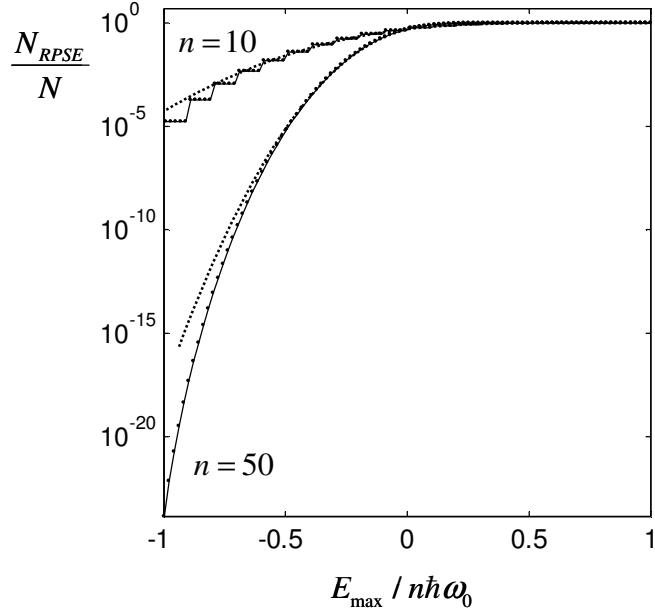

**Fig. 5: Ratio between the dimensions of RPSE active space and of the overall Hilbert space, in a logarithmic scale, as function of the scaled cut-off energy. Dotted lines: eq. (32) resulting from the Gaussian density of states; continuous lines: exact calculations for a system of $n=10$ and of $n=50$ spins, respectively.**

In Fig. 5 we have represented such an asymptotic result for the dimension of the RPSE active space, together with the exact counterpart for systems with $n=10$ and $n=50$ spins and derived by adding the degeneracy eq. (7) as

$$N_{RPSE}(n, E_{max}) = \sum_{i=-n}^{E_{max}/\hbar\omega_0} D(n,i) \qquad (33)$$

Discontinuities of the plotted function are evident for the smaller system ($n=10$). This is a direct consequence of the discrete nature of the energy spectrum, such that no changes of the RPSE properties are observed when the parameter $E_{max}$ is in the domain $i\hbar\omega_0 < E_{max} < (i+1)\hbar\omega_0$ for a given integer $i$, while a step increase is detected when $E_{max} = i\hbar\omega_0$. Of course these discontinuities become less visible for increasing values of $n$. The same kind of behavior will be displayed by other properties in the following Figures.

The comparison made in Figure 5 clearly shows that the Gaussian density of states provides good estimates of $N_{RPSE}$ for positive or nearly vanishing values of $E_{max}$, while significant deviations emerge for negative values of $E_{max}$. This can be understood on a general ground by recalling that one recovers the Gaussian approximation of a given distribution from the parabolic expansion of its logarithm about the maximum. Correspondingly, deviations emerge on the distribution tails far from the maximum. This does not produce significant effects as long as the integral of the density of states eq.(31) includes the maximum (see Fig. 2). On the contrary, the tails have large effects when the integral does not include the maximum, that is for negative $E_{max}$. Therefore, an approximation more accurate than the Gaussian distribution and capable of reproducing correctly the distribution tail is required. Such an approximation will be derived in the following by taking into account that our objective is the calculation in the limit of large $n$ and this justifies the replacement of key parameters with the corresponding leading terms with respect to $n$.

Our starting point is eq. (33) for dimension of the RPSE active space together with the relation eq. (7) for the degeneracy. Notice that when the cut off energy assumes its maximum value, $E_{max} = n\hbar\omega_0$, then from eq. (33) we recover the overall dimension $N(n) = 3^n$ of the Hilbert space for the system

$$N_{RPSE}(n, n\hbar\omega_0) = N(n) \tag{34}$$

In Appendix A it is shown that by scaling the indexes according to the number of spins

$$q_0 = i_0 / n \qquad q = i / n \qquad q_{max} = E_{max} / n\hbar\omega_0 \tag{35}$$

and by replacing the summation on the indexes $i$ and $i_0$ with the corresponding integrals on the $q$ and $q_0$ variables, we get the following relation for the dimension of the RPSE active space

$$N_{RPSE}(n, E_{max}) = n^2 \int_{-1}^{q_{max}} dq\, e^{\hat{d}(n,q)} \tag{36}$$

where the characteristic function $\hat{d}(n,q)$ at the leading (first) order in $n$ reads

$$\hat{d}(n,q) = -n \sum_{s=-1}^{1} \hat{f}_s(q) \ln \hat{f}_s(q) \tag{37}$$

with the three elementary functions $\hat{f}_s(q)$ given by

$$\hat{f}_{\pm 1}(q) = \frac{1 - \alpha(q) \pm q}{2} \qquad \hat{f}_0(q) = \alpha(q) \tag{38}$$

where $\alpha(q) = \dfrac{\sqrt{4 - 3q^2} - 1}{3}$. These functions have the following properties

$$\sum_{s=-1}^{1} \hat{f}_s(q) = 1 \tag{39}$$

$$\hat{f}_0(q)^2 = \hat{f}_{-1}(q) \hat{f}_1(q) \tag{40}$$

By rearranging eq. (37) according to the previously reported properties of the elementary functions, one obtains

$$\hat{d}(n,q)/n = -\ln \hat{f}_0(q) - \frac{q}{2} \ln \frac{\hat{f}_1(q)}{\hat{f}_{-1}(q)} \tag{41}$$

The characteristic function is symmetric with respect to $q$, $\hat{d}(n,-q) = \hat{d}(n,q)$, and reaches its maximum at $q = 0$. With these ingredients we can evaluate the dimension of the RPSE active space according to eq. (36). In order to eliminate the $n^2$ factor, it is convenient to analyze its ratio with respect to the dimension $N(n)$ of the overall Hilbert space, to be specified according to eq. (34)

$$\frac{N_{RPSE}(n, E_{max})}{N(n)} = \frac{N_{RPSE}(n, E_{max})}{N_{RPSE}(n, n\hbar\omega_0)} = \frac{\int_{-1}^{q_{max}} dq\, e^{\hat{d}(n,q)}}{\int_{-1}^{1} dq\, e^{\hat{d}(n,q)}} \tag{42}$$

If $E_{max} > 0$ (and $q_{max} > 0$ as well), then in both integrals of the previous equation we can replace $\hat{d}(n,q)$ with its parabolic expansion about the maximum

$$\hat{d}(n,q) \simeq \hat{d}(n,0) + \frac{\hat{d}''(n,0)}{2} q^2 \tag{43}$$

where $\hat{d}''(n,0) = \partial^2 \hat{d}(n,q)/\partial q^2 \vert_{q=0} < 0$. This leads to a Gaussian profile for the density of states, in strict analogy to eq. (12). In the asymptotic limit $n \to \infty$, by taking into account that according to eq. (41) $\hat{d}''(n,0) \propto n$, a vanishing width is recovered for the Gaussian distribution, which then can be replaced with a Dirac delta. Correspondingly the exponential of the characteristic function can be approximated as

$$e^{\hat{d}(n,q)} = e^{\hat{d}(n,0)} \sqrt{2\pi / \hat{d}''(n,0)}\; \delta(q) \tag{44}$$

and this leads to a unitary value for the ratio of eq.(42)

$$E_{max} > 0: \quad \frac{N_{RPSE}(n, E_{max})}{N(n)} = 1 \tag{45}$$

For $E_{max} < 0$ (and, therefore, for $q_{max} < 0$) we cannot employ eq. (44) to evaluate the integral at the numerator of eq. (42) since the integration domain does not include the maximum of $\hat{d}(n,q)$ (of course, we can continue to use it for the integral at the denominator). Then we must introduce a different kind of approximation. By taking into account that, for $q < 0$, $\hat{d}(n,q)$ is an increasing function of $q$ with a slope proportional to the size $n$ of the system, in the asymptotic limit $n \to \infty$ we can replace $\hat{d}(n,q)$ with its linear expansion about the upper integral boundary:

$$\hat{d}(n,q) \simeq \hat{d}(n,q_{max}) + \hat{d}'(n,q_{man})(q - q_{max}) \tag{46}$$

where the first derivative of the characteristic function is explicitly given as

$$\hat{d}'(n,q) \equiv \frac{\partial \hat{d}(n,q)}{\partial q} = -n \sum_{s=-1}^{1} \frac{\partial \hat{f}_s(q)}{\partial q} \ln \hat{f}_s(q) = -\frac{n}{2} \ln \frac{\hat{f}_1(q)}{\hat{f}_{-1}(q)} \tag{47}$$

where to obtain the r.h.s. the derivatives of the elementary functions eq. (38) have been explicitly calculated. By extending the lower integral boundary to $-\infty$, and by evaluating the numerator according to eq. (46), we get the following asymptotic relation for the dimension of the RPSE active space

$$\frac{N_{RPSE}(n, E_{max})}{N(n)} = \frac{\sqrt{\hat{d}''(n,0)/2\pi}}{\hat{d}'(n,q_{max})} \exp\{\hat{d}(n,q_{max}) - \hat{d}(n,0)\} \qquad (48)$$

Finally, by considering the logarithm of the previous ratio, we extract the leading (linear) contributions with respect to $n$

$$E_{max} < 0: \quad \ln \frac{N_{RPSE}(n, E_{max})}{N(n)} = \hat{d}(n, q_{max}) - \hat{d}(n, 0) = -n \sum_{s=-1}^{1} \hat{f}_s(q_{max}) \ln \hat{f}_s(q_{max}) - n \ln 3 \qquad (49)$$

where the derivatives of $\hat{d}(n,q)$ entering in eq. (48) do not appear because they contribute like $\ln n$. Notice that the two asymptotic approximations eq. (49) and eq. (45) match exactly at $E_{max} = 0$. In order to show how such an asymptotic profile is reached by increasing the system size, in Fig. 6 we have represented the logarithm of $N_{RPSE}$ derived from eq. (49) and scaled by the number of spins as a function of $E_{max}$, together with the corresponding exact values of finite systems.

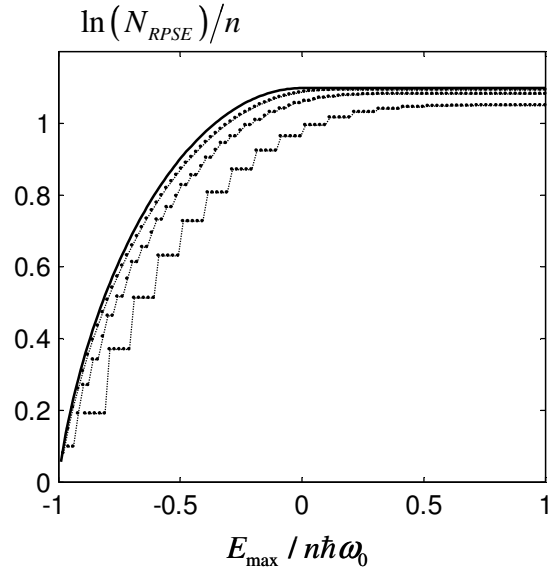

**Figure 6: Logarithm of the dimensions of RPSE active space, divided by the number of spin, as a function of the energy cut off. Continuous line: asymptotic dependence eq. (49) and eq. (45); dotted lines: exact results for finite systems approaching the asymptotic profile for an increasing number of spins,** $n = 10, 30, 100$.

Having characterized the dimension of the RPSE active space, now we can calculate the thermodynamic properties on the basis of the RPSE statistics. The probability distribution on populations is uniform on the $N$ simplex defined by the normalization constraints and can be approximated by a factorized probability distribution derived in Ref. 5, namely

$$p(P) \simeq \prod_{k=1}^{N_{RPSE}} w(P_k) \qquad w(P_k) = N_{RPSE} e^{-N_{RPSE} P_k} \tag{50}$$

each population $P_k$ being considered in the entire positive real axis. Notice that in the RPSE all the populations are equivalent in the meaning that they are identically distributed, and the moments of their distribution are given by

$$\langle P_k^m \rangle = \frac{m!}{N_{RPSE}^m} \tag{51}$$

As already mentioned in I, the internal energy of the Random Pure State Ensemble is not an independent parameter as in the FEEE, but it is identified with the typical value of the expectation energy in the ensemble. Thus, the definition of the internal energy as the average value of the expectation energy, $U = \langle E \rangle$, is meaningful only as long as the ensemble distribution of the expectation energy $E = \sum_k P_k E_k$ is a function peaked at a typical value. In order to verify this condition let us calculate its average and its variance within the population distribution

$$\langle E \rangle = \sum_{k=1}^{N_{RPSE}} \langle P_k \rangle E_k = \frac{1}{N_{RPSE}} \sum_{k=1}^{N_{RPSE}} E_k \tag{52}$$

$$\sigma_E^2 = \langle E^2 \rangle - \langle E \rangle^2 = \sum_{k=1}^{N_{RPSE}} E_k^2 \left( \langle P_k^2 \rangle - \langle P_k \rangle^2 \right) = \frac{1}{N_{RPSE}^2} \sum_{k=1}^{N_{RPSE}} E_k^2 \tag{53}$$

where for the r.h.s eq. (51) has been used. First, let us analyze in more detail the internal energy which, by employing the degeneracy eq. (7) together with eq. (33) for the dimension of the RPSE active space, can be specified as

$$U(n, E_{max}) / \hbar \omega_0 = \frac{\sum_{n=-n}^{E_{max}/\hbar\omega_0} i D(n,i)}{\sum_{n=-n}^{E_{max}/\hbar\omega_0} D(n,i)} \tag{54}$$

where we have explicitly denoted the dependence of the internal energy on the number of spins, and on the energy cut off. Such an equation allows the efficient calculation of the internal energy for finite systems. When the asymptotic behaviour for $n \to \infty$ of the energy is required, then it is convenient to perform the same change of variables leading to eq. (36), so obtaining

$$U(n, E_{\max})/n\hbar\omega_0 = \frac{\int_{-1}^{q_{\max}} dq\, e^{\hat{d}(n,q)} q}{\int_{-1}^{q_{\max}} dq\, e^{\hat{d}(n,q)}} \tag{55}$$

In order to derive the asymptotic value of the internal energy for $E_{\max} > 0$, one can evaluate the integrals of the previous equation according to the approximation eq. (44) so obtaining

$$E_{\max} > 0: \quad U(n, E_{\max}) = 0 \tag{56}$$

For $E_{\max} < 0$, the linear expansion eq. (46) for the characteristic function has to be employed with the following result

$$U(n, E_{\max})/n\hbar\omega_0 = q_{\max} + \frac{1}{\hat{d}'(n, q_{\max})} \tag{57}$$

In the asymptotic limit $n \to \infty$, the last term at the r.h.s. can be neglected since $\hat{d}'(n,q) \propto n$, so that

$$E_{\max} < 0: \quad U(n, E_{\max}) = E_{\max} \tag{58}$$

In Fig. 7 we have reported these asymptotic equations (56) and (58), together with the exact values eq. (54) of the internal energy of finite systems, in order to show the convergence to the asymptotic regime by increasing the number of spins. To demonstrate that the internal energy, eqs. (56) and (58), is the typical energy of the RPSE we have to consider its variance. For the model at hand, eq. (53) can be written as

$$(\sigma_E/\hbar\omega_0)^2 = \frac{1}{N_{RPSE}^2} \sum_{n=-n}^{E_{\max}/\hbar\omega_0} i^2 D(n,i) \tag{59}$$

The asymptotic calculation of $\sigma_E$ is a bit more involved and we report here only the final result: $\sigma_E^2 \propto n^2/N_{RPSE}$. Since the range of the possible energies, $\Delta E = 2n\hbar\omega_0$, increases linearly with the number of components $n$ the typicality of the expectation energy is assured, i.e. $\sigma_E/\Delta E \propto 1/\sqrt{N_{RPSE}}$.

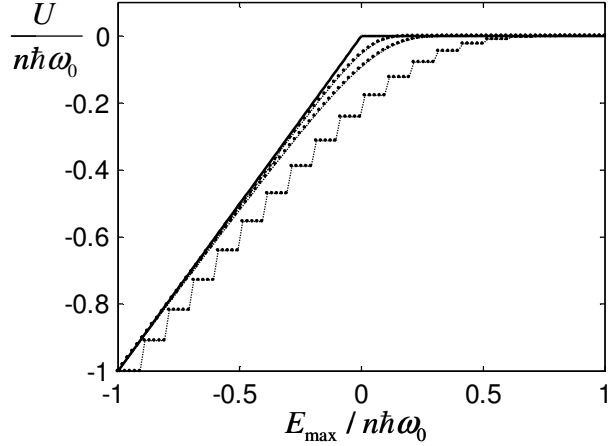

**Fig. 7: RPSE internal energy a function of the energy cut off. Continuous line: asymptotic behaviour described by eqs. (56) and (58). The dotted lines represent the result for finite system approaching the asymptotic profile for an increasing number of spins,** $n = 10, 50, 150$.

Let us now examine the entropy to be evaluated according to eq. (15) by means of the RPSE distribution described by eq. (50), which leads to the following results for its average value

$$\langle S \rangle = k_B \ln N_{RPSE} - (1-\gamma) \tag{60}$$

where $\gamma$ is the Euler constant, and with eq. (33) providing the value for $N_{RPSE}$. We do not analyze here the entropy variance $\sigma_S$ since in Ref. 5 we have already shown that $\sigma_S / \Delta S \propto 1/\sqrt{N_{RPSE}}$ implying that typicality is recovered for moderately large systems. As long as such a property can be safely taken for granted, in the following the average (typical) entropy will be recalled simply as the system's entropy and denoted as $S$ without reporting the bracket denoting the RPSE average for the sake of a more compact notation. Moreover, we do not include any specific plot for the entropy since, according to eq. (60), its behaviour can be assimilated to that of $\ln(N_{RPSE})$ already analyzed in Fig. 6. In order to show that the RPSE entropy has the correct thermodynamic behaviour, its asymptotic form for $n \to \infty$ has to be considered by inserting approximations eqs. (45) and (49) into eq. (60). By retaining only the leading (linear) contribution with respect to $n$, we get

$$\begin{aligned} E_{\max} < 0: \quad & S(n, E_{\max}) = k_B \hat{d}(n, E_{\max} / n\hbar\omega_0) \\ E_{\max} > 0: \quad & S(n, E_{\max}) = n k_B \ln 3 \end{aligned} \tag{61}$$

For $E_{\max} < 0$, by taking into account that according to eq. (58) the internal energy and the energy cut off coincide, one derives the following equation of state for the entropy as a function of the internal energy

$$S = k_B \hat{d}(n, U / n\hbar\omega_0) = -n k_B \sum_{s=-1}^{1} \hat{f}_s(U / n\hbar\omega_0) \ln \hat{f}_s(U / n\hbar\omega_0) \tag{62}$$

For $E_{max} > 0$, both $S$ and $U$ are constant (that is, they are independent of $E_{max}$) and, therefore, they identify a unique thermodynamic state with the same entropy of the limit of eq. (62) for $U \to 0^-$ (i.e., for $E_{max} \to 0^-$). Therefore eq. (62) supplies the full equation of state for the entropy with an extensive character since a given internal energy per component (i.e., $U/n$) determines the corresponding entropy per component $S/n$. Furthermore, such an equation of states allows the direct determination of temperature according to the thermodynamic derivative $1/T = dS/dU$, so deriving

$$\frac{\hbar\omega_0}{k_B T} = \frac{\hat{d}'(n, U/n\hbar\omega_0)}{n} \qquad (63)$$

where $\hat{d}'(n,q)$ is explicitly given in eq. (47).

In Fig. 8 we have represented the profile of the inverse temperature against the internal energy (with suitable scaling factors). On the basis of this result, we can conclude that RPSE supplies an equation of state for the entropy in agreement with the well known thermodynamic behaviour, since the increase of the internal energy produces always a positive increment of the temperature, which corresponds to a convex profile of the state function $S(U)$. Notice also that $1/T$ vanishes for $U \to 0$ in correspondence of the thermodynamic state with infinite temperature.

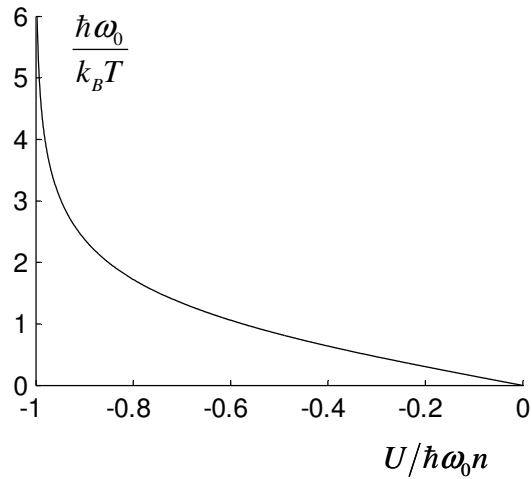

**Fig. 8: Inverse temperature as function of the internal energy from RPSE**

To complete our analysis, it remains to investigate the properties of the typical equilibrium state of a subsystem, i.e. the equilibrium average $\bar{\mu}$ of the reduced density matrix of a single spin. According to the relation eq. (24), the elements of $\bar{\mu}$ are given by the sum of identically distributed random variables, i.e. the global population whose distribution is given by eq. (50). Then, the RPSE average of the reduced density matrix elements can be written as

$$\langle \bar{\mu}_{ss} \rangle = \sum_e{}' \langle P_{se} \rangle \qquad (64)$$

where, in order to satisfy the constraint on the energy cut-off $E_{max}$, and by taking into account that the subsystem contribution to the energy is specified as $s\hbar\omega_0$, the summation $\sum_e'$ on the environment states should be confined to those environment configurations which contribute to the energy at most as $E_{max} - s\hbar\omega_0$. The number of these configurations is $N_{RPSE}(n-1, E_{max} - s\hbar\omega_0)$ as derived by considering the environment alone, that is, the system of $(n-1)$ spins. Then, by taking into account that for the RPSE the average populations are identical and equal to $1/N_{RPSE}(n, E_{max})$, we get the following relation for the reduced density matrix elements

$$\langle \bar{\mu}_{ss} \rangle = \frac{N_{RPSE}(n-1, E_{max} - s\hbar\omega_0)}{N_{RPSE}(n, E_{max})} \tag{65}$$

The variance of these quantities can be evaluated by applying the Central Limit Theorem to the linear combination of population in eq. (24), so obtaining that $\sigma_{\bar{\mu}_{ss}} \propto \langle \bar{\mu}_{ss} \rangle / N_{RPSE}(n-1, E_{max} - s\hbar\omega_0)$ implying typicality for a large enough environment. In Fig. 9 we have represented the dependence of the RPSE average eq. (65) of the reduced density matrix elements on the scaled cut-off energy $E_{max}/n\hbar\omega_0$ for some finite systems, with the dimension $N_{RPSE}$ evaluated according to eq. (33).

In order to provide an asymptotic estimate of the RPSE average of the reduced density matrix elements for a thermodynamic state with a given finite temperature, we can employ eq.(49) to specify $N_{RPSE}$ for $E_{max} < 0$, so obtaining the following relation for the asymptotic limit

$$\lim_{n\to\infty} \ln\langle \bar{\mu}_{ss} \rangle = \lim_{n\to\infty} \left[ \hat{d}(n-1, q'_{max}) - \hat{d}(n, q_{max}) \right] \tag{66}$$

where

$$q_{max} = \frac{E_{max}}{n\hbar\omega_0} \qquad q'_{max} = \frac{E_{max} - s\hbar\omega_0}{(n-1)\hbar\omega_0} = \frac{nq_{max} - s}{n-1} \tag{67}$$

It should be stressed that, as long as the temperature is fixed, $q_{max}$ is a constant, while $q'_{max}$ has an explicit dependence on the number $n$ of spins. In order to evaluate the previous limit, it is convenient to replace $n$ with $1/x$ and to rewrite the argument of the limit as a ratio

$$\lim_{n\to\infty} \ln\langle \bar{\mu}_{ss} \rangle = \lim_{x\to 0} \frac{x\hat{d}(x^{-1}-1, (q_{max}-sx)/(1-x)) - x\hat{d}(x^{-1}, q_{max})}{x} \tag{68}$$

so that we can employ de l'Hopital rule

$$\lim_{n\to\infty} \ln\langle \bar{\mu}_{ss} \rangle = \lim_{x\to 0} \frac{\partial}{\partial x}\left[ x\hat{d}(x^{-1}-1, (q_{max}-sx)/(1-x)) - x\hat{d}(x^{-1}, q_{max}) \right] \tag{69}$$

By inserting the explicit form eq. (41) for the characteristic function, and by noting also that $x\hat{d}(x^{-1}, q_{max}) = \hat{d}(n, q_{max})/n$ is independent of $x$, we get finally

$$\lim_{n\to\infty} \ln\langle \bar{\mu}_{ss} \rangle = -\frac{\hat{d}(n, q_{max})}{n} + \frac{\hat{d}'(n, q_{max})}{n}(q_{max} - s) = \ln\hat{f}_0(n, q_{max}) - s\frac{\hat{d}'(n, q_{max})}{n} \tag{70}$$

where the explicit forms of the characteristic function $\hat{d}(n, q_{max})$, eq.(41), and its first derivative, eq. (47), have been employed to derive the r.h.s.. Such a result, by taking into account the previously derived relation eq. (63) for the thermodynamic temperature, implies the Boltzmann canonical form eq. (26) for the typical reduced density matrix.

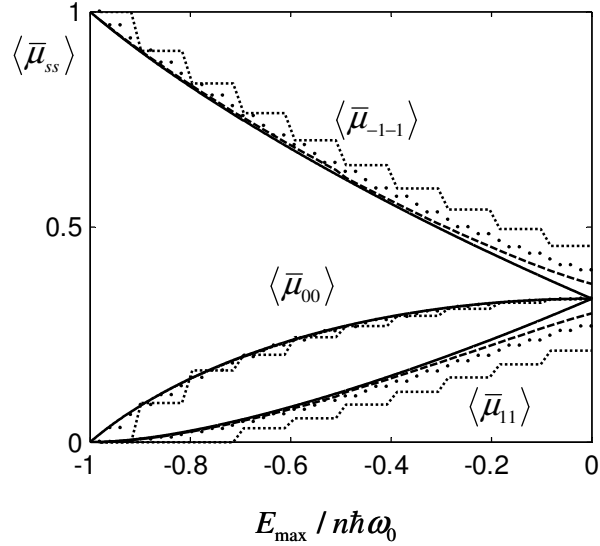

**Fig. 9: Dependence of the RPSE average of the reduced density matrix elements on the scaled cut-off energy. The asymptotic function which describes the dependence of the typical values on the energy eq. (71) is reported as continuous lines. The other lines represent the result for finite system approaching the asymptotic profile for an increasing number of spins,** $n = 10, 30, 100$.

Interestingly, eq. (70) implies the following direct relation between the average of the reduced density matrix elements and the elementary functions previously introduced

$$n \to \infty: \qquad \langle \bar{\mu}_{ss} \rangle = \hat{f}_s(q_{max}) \tag{71}$$

For $s = 0$, such a relation is a direct consequence of eq. (70), while for $s = 1$, by specifying the derivative of the characteristic function according to eq. (47), one gets

$$\langle \bar{\mu}_{11} \rangle = \hat{f}_0(q_{max}) \exp\left[-\hat{d}(n, q_{max})/n\right] = \sqrt{\hat{f}_0(q_{max})^2 \hat{f}_1(q_{max}) / \hat{f}_{-1}(q_{max})} = \hat{f}_1(q_{max}) \tag{72}$$

with condition eq. (40) leading to the r.h.s.. Similarly one proceeds for the $s = -1$ case. Notice that the normalization of the reduced density matrix

$$\sum_{s=-1}^{1} \langle \bar{\mu}_{ss} \rangle = 1 \tag{73}$$

is assured also for the asymptotic form eq. (71) as a consequence of the condition eq. (39) for the elementary functions. In Fig. 9, together with the results for finite systems, we have also represented the asymptotic result eq. (71) in order to provide a direct evidence of the convergence for an increasing number of spins. Because of the relation (71), the entropy of the isolated system eq. (62) divided for the number of spins, assumes the form corresponding to the canonical statistics

$$n \to \infty: \qquad S/n = -k_B \sum_{s=-1}^{1} \langle \bar{\mu}_{ss} \rangle \ln \langle \bar{\mu}_{ss} \rangle \qquad (74)$$

That is, also the overall entropy has a canonical form since we are dealing with an ideal system (i.e., non interacting components). This represents the RPSE counterpart of a similar statement of standard statistical thermodynamics about the equivalence of the canonical entropy and the microcanonical entropy.

In conclusion, the RPSE satisfies all the requirements invoked in part I for generating on a quantum mechanical basis the thermodynamic behaviour when dealing with large enough spin systems.

The results of this section can be compared with the *Canonical Typicality* proved by Goldstein et al. [10] who have considered as active space the portion of the Hilbert space spanned by the set of Hamiltonian eigenfunctions whose eigenvalues belong to a certain thin slice of the spectrum, i.e. $E_k \in [E_{\min}, E_{\max}]$ with $E_{\max} - E_{\min} \ll E_{\max}$. The RPSE is an extension of such an ensemble of pure states since $E_{\min} \to -\infty$. However, canonical typicality of Ref [10] is proved for the reduced density matrix and not for its asymptotic time average as we have done here. Thus, in our framework the same result can be reached in two steps: firstly by demonstrating the typicality of the asymptotic time average in correspondence to the canonical form and then by noting that temporal fluctuations around such an average are small in an appropriate scale. This is easily proved as follows: as already stressed in I the PSD derived on the basis of the dynamics of the isolated system provides the fluctuation amplitude as a specific function of the population set. For a generic observable A one has (eq. 17 of I)

$$\overline{\Delta a^2} = \sum_{k' \neq k}^{N_{RPSE}} A_{kk'} A_{kk'}^* P_k P_{k'} \qquad (75)$$

Its average value in the RPSE reads

$$\left\langle \overline{\Delta a^2} \right\rangle = \frac{1}{N_{RPSE}^2} \sum_{k \neq k'}^{N_{RPSE}} A_{kk'} A_{kk'}^* \leq \frac{\mathrm{Tr}\{\wp A A^\dagger\}}{N_{RPSE}^2} \qquad (76)$$

where we have introduced the projection operator onto the active space of the RPSE, that is $\wp = \sum_{k}^{N_{RPSE}} |k\rangle\langle k|$. Notice that the trace on the r.h.s of eq.(76) can be decomposed into the trace over the subsystem S and the environment E, $\mathrm{Tr}(...) = \mathrm{Tr}_S \mathrm{Tr}_E(...)$. If the operator A describes a subsystem property, eq. (76) becomes

$$\left\langle \overline{\Delta a^2} \right\rangle \leq \frac{\sum_{ss'}^{N_S} (AA^\dagger)_{ss'} \sum_{e}^{N_E} \wp_{s'e,se}}{N_{RPSE}^2} \qquad (77)$$

where $N_S$ and $N_E$ are the dimension of the Hilbert spaces for the subsystem ($S$) and the environment ($E$), respectively. For the ideal system of $n$ spins here considered, by taking into account eq. (65), we get the relation

$$\sum_{e}^{N_E} \wp_{s'e,se} = \delta_{ss'} N_{RPSE}(n-1, E_{max} - s\hbar\omega_0) = N_{RPSE}(n, E_{max})\langle \bar{\mu}_{ss} \rangle \qquad (78)$$

which implies the following upper bound for the average fluctuation amplitude

$$\overline{\langle \Delta a^2 \rangle} \leq \frac{\mathrm{Tr}_S\{AA^\dagger \langle \bar{\mu} \rangle\}}{N_{RPSE}} \qquad (79)$$

Since $N_{RPSE}$ increases exponentially with the number of components in the global system, the inequality in eq. (79) implies typicality in the same meaning of ref. [10] for any subsystem properties, elements of the reduced density matrix included.

Notice that the active space of the RPSE is not limited to a "macroscopically small" energy interval as considered by Goldstain et al.. On the contrary all the energy eigenstates with energy lower than the internal energy can participate to the pure state. However, due to the strong increasing of the energy density for a thermodynamic system with finite temperature, the two ensembles lead to similar results. Thus, to recover canonical typicality for the subsystem equilibrium state it is not necessary to restrict the possible wavefunctions to superposition of energy eigenstates which belong to a "thin" energy slice because more general superposition states leads to the same conclusions.

## V- DISCUSSION AND CONCLUSIONS

The theoretical framework presented in this work is aimed to clarify the connections between a pure quantum mechanical description of a composite quantum system in a pure state and its characterization in terms of thermodynamic quantities. These connections are naturally of statistical nature, and the logical steps to account for the emergence of thermodynamic properties are the following:

1) verify the existence of a typical value of thermodynamic functions in an ensemble chosen for the statistical sampling of the populations which, together with the phases, parameterize the system wavefunction.

2) verify that typical values in a particular ensemble behave as thermodynamics requires.

Moreover, in order to recover the standard results of statistical thermodynamics, we include a further requirement

3) verify the typicality and the canonical form of the reduced density matrix of the subsystem.

The statistical character of such a theoretical framework entails that it makes predictions only when, and to the extent that, it leads to sharp distributions on the "macroscopic" observables despite the existence of a wide range of possible "microstate", i.e. different sets of populations. In our case the "macroscopic" observable is the value of the entropy, or the elements of the reduced density matrix of a subsystem. The emergence of a "typical" value for these properties is thus necessary for the mere existence of a statistical theory for them. Evidently the sharp distribution on the macroscopic observables can emerge only if each of the overwhelming majority of the microscopic states having an appreciable weight in the considered ensemble is characterized by nearly the same value of the macroscopic functions. In this spirit we can understand the emergence of thermodynamic properties by considering quantum pure states. If the theory predicts values of macroscopic functions which do not agree with thermodynamics, as in the case of the FEEE, then it is reasonable to conclude that the possible states in the ensemble are not correctly weighted. On the other hand we have demonstrated that typical values of the investigated properties in the RPSE satisfy all the requirements necessary for a meaningful thermodynamical characterization of the isolated system. Notably, the consistency with thermodynamics is not restricted to pure states which are superposition of energy eigenstates corresponding to eigenenergies belonging to a "thin energy shell" which is usually associated to the quantum microcanonical statistics, but is fairly more general. Moreover two not obvious properties have been rigorously proven to hold for the model system at hand: i) the correspondence of the global temperature derived from the entropy equation of state $S(U)$ and the local one which appears in the canonical equilibrium state of a subsystem and ii) the equivalence between the RPSE typical entropy and the canonical entropy at the same internal energy. In the present paper the analysis is made for a model system composed of identical spin with $J=1$ for illustrative purposes, however a straightforward generalization to systems with arbitrary energy spectrum is possible and it will be presented elsewhere.

To conclude the presentation of part I and part II of this work, we would like to review the methodological choices assumed by us. The analysis is based on two main tools: The Pure State Distribution and the statistical ensemble of pure states. The former is fully characterized on the basis of the dynamics of the quantum pure state and allows the determination of the entire distribution of any property of the quantum state; it does not lead to any problematic issue. The statistical ensemble of pure state is also a necessary ingredient, as long as the populations are constants of the motion and there is no a priori motivation for the selection of them. Different statistical ensembles can be proposed and some criterion of choice has to be introduced. We have considered the agreement of the predictions resulting from ensemble statistics with the macroscopic thermodynamics on the one hand, and with statistical thermodynamics, in relation to the canonical state of a subsystem, on the other hand. This has allowed us to show that FEEE is not an adequate ensemble, while RPSE fulfils all the invoked requirements. Then the following question arises: is RPSE the only allowed ensemble? The answer is certainly negative, as long as one can easily devise other ensembles satisfying the same requirements. For instance, one can consider a variant of the RPSE by regarding as active space the Hilbert subspace spanned by the Hamiltonian eigenvectors within the energy shell $E_{min} < E_k \leq E_{max}$. In this case, however, the typical entropy $\langle S \rangle$ and the internal energy, i.e., the ensemble average of the expectation energy, $U = \langle E \rangle$, become functions of both parameters $E_{min}$ and $E_{max}$. Therefore, in order to recover the entropy equation of state, $\langle S \rangle(U)$, from the elimination of one parametric dependence one has to introduce a constraint between these parameters, say that $E_{min}$ is a given fraction of $E_{max}$. Still, also in this case, one would obtain predictions in agreement with macroscopic thermodynamics and with the canonical statistics of a subsystem. Therefore, if these are the only requirements for the ensemble, different choices are possible. In such a framework, however, we think that our proposal of RPSE is the most convenient because of its very simple (mono-parametric) definition and because subjective choices, like for the ratio $E_{min}/E_{max}$, are not required. On the other hand, the following question naturally arises: can one introduce further requirement of a different nature for the choice of the statistical ensemble? If the answer is positive, then more stringent criteria would be available for the selection of the statistical ensemble. This could be the objective of future challenging researches.

Lastly, we would like to emphasize that the statistical theory we have presented is intrinsically defined for finite systems, even if in the applications the asymptotic properties in the limit of large systems have been often considered. Therefore, it could be well employed to characterize the equilibrium properties of those finite systems which, in recent years, have been analyzed in relation to the dynamical and the relaxation behaviour [11].

### Acknowledgements

The authors acknowledge the support by Univesità degli Studi di Padova through 60% grants.### Appendix: Evaluation of the characteristic function $\hat{d}(n,q)$

The starting point is eq. (33) for the dimension of the RPSE active space, with the degeneracy evaluated according to eq. (7). After substituting the summations on the indices $i$ and $i_0$ with the integrals on parameters $q$ and $q_0$ given in eq. (35), we get eq. (36) with

$$e^{\hat{d}(n,q)} \equiv \int_0^{\text{Inf}(1+q,1-q)} dq_0 \, e^{d(n,q,q_0)} \tag{A1}$$

where $d(n,q,q_0)$ is the logarithm of the ratio of factorials in eq. (7)

$$d(n,q,q_0) \equiv \ln \frac{n!}{[(n-i_0-i)/2]! \, i_0! \, [(n-i_0+i)/2]!} \bigg|_{i=nq, i_0=nq_0} \tag{A2}$$

By means of the Stirling approximation, $m! \simeq \sqrt{2\pi} e^{-m-1}(m+1)^{m+1/2}$, and by retaining only the leading (linear) terms with respect to the parameter $n$, we obtain the following asymptotic form of the function defined by eq. (A2)

$$d(n,q,q_0) = -n \sum_{s=-1}^{1} f_s(q,q_0) \ln f_s(q,q_0) \tag{A3}$$

with the three elementary functions $f_s(q,q_0)$

$$f_{\pm 1}(q,q_0) = \frac{1-q_0 \pm q}{2} \qquad f_0(q,q_0) = q_0 \tag{A4}$$

decomposing the unity

$$\sum_{s=-1}^{1} f_s(q,q_0) = 1 \tag{A5}$$

In order to derive a suitable approximation for the integral eq. (A1) we locate the maximum of function $d(n,q,q_0)$ eq. (A3) with respect to $q_0$ from its derivative

$$\frac{\partial}{\partial q_0} d(n,q,q_0) = -n \sum_{s=-1}^{1} \frac{\partial f_s(q,q_0)}{\partial q_0} \ln f_s(q,q_0) = -n \ln \frac{f_0(q,q_0)^2}{f_{-1}(q,q_0)f_1(q,q_0)} \tag{A6}$$

where we have taken into account the explicit form eq. (A4) of the elementary function $f_s(q,q_0)$. Thus, the maximum $\alpha(q)$ is located in the correspondence of the positive root of the following equation

$$\left. \frac{f_0(q,q_0)^2}{f_{-1}(q,q_0)f_1(q,q_0)} \right|_{q_0=\alpha(q)} = 1 \tag{A7}$$

which reads explicitly

$$\alpha(q) = \frac{\sqrt{4-3q^2}-1}{3} \tag{A8}$$

In Fig. 10, together with the domain of existence for the function $d(n,q,q_0)$, we have drawn the position $\alpha(q)$ of the maximum. It is evident that the integration domain includes always the maximum location, with only the exception of the extreme cases for $q = \pm 1$ which, however, are not relevant in determining the properties like the dimension of the RPSE active space. This suggests that function $d(n,q,q_0)$ could be substituted by its parabolic expansion about the maximum, that is

$$d(n,q,q_0) \simeq d(n,q,\alpha(q)) + K(n,q)[q_0 - \alpha(q)]^2 / 2 \tag{A9}$$

where

$$K(n,q) \equiv \left. \frac{\partial^2 d(n,q,q_0)}{\partial q_0^2} \right|_{q_0=\alpha(q)} = -2n \frac{1-\alpha(q)-q^2}{\alpha(q)\{[1-\alpha(q)]^2 - q^2\}} \tag{A10}$$

As long as the curvature $K(n,q)$ is negative, the integrand of eq. (A1) has a bell shaped profile with a width, according to eq. (A10), proportional to $1/\sqrt{n}$. Correspondingly, in the asymptotic limit $n \to \infty$, the integrand of eq. (A1) becomes a very peaked function, and this justifies the use of the parabolic expansion eq. (A9) and the extension to $\pm\infty$ of the integration boundaries as well. In conclusion the following approximation is recovered for the characteristic function

$$\hat{d}(n,q) \simeq \ln \int_{-\infty}^{\infty} dq_0 \exp\left\{ d(n,q,\alpha(q)) + K(n,q)[q_0 - \alpha(q)]^2 / 2 \right\} =$$
$$= d(n,q,\alpha(q)) - \frac{1}{2}\ln\left| K(n,q)/2\pi \right| \tag{A11}$$

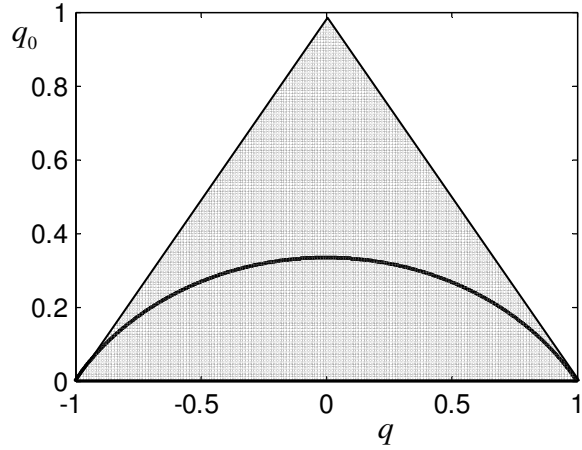

**Fig. 10: Plot of the location $\alpha(q)$ of $d(n,q,q_0)$ maximum with respect to $q_0$ (continuous line) within the existence domain of the function (grey area)**

By retaining only the leading (linear) terms with respect to $n$, the following simple relation is found for the characteristic function for the degeneracy

$$\hat{d}(n,q) = d(n,q,\alpha(q)) = -n \sum_{s=-1}^{1} \hat{f}_s(q) \ln \hat{f}_s(q) \tag{A12}$$

where the new elementary functions $\hat{f}_s(q)$ are the functions defined in (A4) evaluated at $q_0 = \alpha(q)$

$$\hat{f}_s(q) \equiv f_s(q, \alpha(q)) \tag{A13}$$

Their explicit form is reported in eq. (38) of the main text, together with properties eq. (39) and eq. (40) deriving from eq. (A5) and eq. (A7).

Notice that the characteristic function $\hat{d}(n,q)$ determines the density of states $g(E)$

$$g(E) = n\hbar\omega_0 e^{\hat{d}(n, E/n\hbar\omega_0)} \tag{A.14}$$

as one can derive by comparing eq. (31) with eq.(36).